\documentclass[acmtog]{acmart}

\acmJournal{TOG}

\setcopyright{none}
\setcitestyle{square}

\received{April 2017}
\received[final version]{XX 2017}
\received[accepted]{XX 2017}

\usepackage{amsmath,amssymb,amsthm}
\usepackage{bm}
\usepackage{booktabs}
\usepackage[ruled,linesnumbered,vlined,algo2e]{algorithm2e}
\usepackage{graphicx}
\usepackage{epstopdf}
\usepackage{rotating}
\usepackage{multirow}
\usepackage[percent]{overpic}
\usepackage[outline]{contour}
\usepackage{xifthen}
\usepackage[export]{adjustbox}
\contourlength{0.1em}
\definecolor{darkred}{rgb}{0.6, 0.0, 0.1}
\graphicspath{{figures/}}
\usepackage{graphics}
\usepackage{pgf}
\usepackage{mathrsfs}
\usepackage[pstarrows]{pict2e}

\definecolor{WAJcolor}{rgb}{0.0,0.6,0.0}
\definecolor{WKJcolor}{rgb}{0.8,0.0,0.4}
\definecolor{BBcolor}{rgb}{0.0,0.6,0.8}
\definecolor{JNcolor}{rgb}{0.8,0.6,0.0}

\def\commentType{0}

\ifnum\commentType=0
	\newcommand{\customComment}[3]{}
	\newcommand{\customTODO}[3]{}
\fi
\ifnum\commentType=1
	\newcommand{\customComment}[3]{\textcolor{#2}{\textsl{#1: #3}}}
	\newcommand{\customTODO}[3]{\textcolor{#2}{\textsl{#1 TODO: #3}}}
\fi
\ifnum\commentType=2
	\usepackage{pdfcomment}
	\newcommand{\customComment}[3]{\pdfcomment[icon=Comment,opacity=0.5,color=#2,author=#1]{#3}}
	\newcommand{\customTODO}[3]{\pdfcomment[icon=Note,opacity=0.5,color=#2,author=#1]{#3}}
\fi
\ifnum\commentType=3
	\usepackage{pdfcomment} 
	\usepackage{todonotes} 
	\usepackage{silence}
	\newcommand{\customComment}[3]{\todo[color=#2!40,size=\small]{\textbf{#1:} #3}}
	\newcommand{\customTODO}[3]{\todo[inline,color=#2!40,size=\small]{\textbf{#1:} #3}}
\fi

\newcommand{\secref}[1]{Section~\ref{#1}}
\newcommand{\equref}[1]{Equation~\eqref{#1}}
\newcommand{\figref}[1]{Figure~\ref{#1}}
\newcommand{\tabref}[1]{Table~\ref{#1}}

\newcommand{\punct}[1]{\,#1}

\newcommand{\diff}[1]{\mathrm{d}#1}

\newcommand{\Measurement}{I}

\newcommand{\MContrib}{f}
\newcommand{\MContribPSS}{f}
\newcommand{\MImportance}{f}
\newcommand{\MImportancePSS}{f}
\newcommand{\MPSSTarget}{C}
\newcommand{\MPSSTargetImportance}{C}
\newcommand{\MPSTarget}{C}
\newcommand{\MPSTargetImportance}{C}

\newcommand{\ReconFilter}[1]{h_{#1}}

\newcommand{\dArea}[1]{\diff{A}\ifthenelse{\isempty{#1}}{}{(#1)}}
\newcommand{\dSAngle}[1]{\diff{\sigma}\ifthenelse{\isempty{#1}}{}{(#1)}}
\newcommand{\dPSAngle}[2]{\diff{\sigma^\perp}\ifthenelse{\isempty{#1}}{}{(#1, #2)}}
\newcommand{\dPath}[1]{\diff{\mu}\ifthenelse{\isempty{#1}}{}{(#1)}}

\newcommand{\PathSpace}{\mathcal{P}}
\newcommand{\PSSpace}{\mathcal{U}}
\newcommand{\ExtendedPSSpace}{\mathscr{C}}
\newcommand{\PSSpaceSize}{o}
\newcommand{\stCount}{q}

\newcommand{\pdf}{p}
\newcommand{\pdfPSS}{p}

\newcommand{\xpoint}{\mathbf{x}}
\newcommand{\ypoint}{\mathbf{y}}

\newcommand{\xbar}{\bar{\xpoint}}
\newcommand{\ybar}{\bar{\ypoint}}

\newcommand{\xbarAt}[1]{\xpoint_{#1}}
\newcommand{\ybarAt}[1]{\ypoint_{#1}}

\newcommand{\upoint}{\mathbf{u}}
\newcommand{\vpoint}{\mathbf{v}}
\newcommand{\ubar}{\bar{\upoint}}
\newcommand{\vbar}{\bar{\vpoint}}
\newcommand{\ubarAt}[1]{\upoint_{#1}}
\newcommand{\vbarAt}[1]{\vpoint_{#1}}

\newcommand{\tightoverset}[2]{%
  \mathop{#2}\limits^{\vbox to -.8ex{\kern-0.75ex\hbox{$#1$}\vss}}}

\newcommand{\auxScalar}{\gamma}
\newcommand{\aux}{\bm{\auxScalar}}
\newcommand{\auxbar}{\bar{\aux}}
\newcommand{\auxbarAt}[1]{\aux_{#1}}

\newcommand{\misWeight}{w}
\newcommand{\misWeightPSS}{w}

\newcommand{\StationaryDistribution}{C}

\newcommand{\ProposalDistribution}{T}

\newcommand{\CurrentState}{\mathbf{x}}

\newcommand{\NextState}{\mathbf{y}}
\newcommand{\CurrentStatePSS}{\mathbf{\bar u}}

\newcommand{\CurrentStateExPSS}{\mathbf{\hat u}}
\newcommand{\NextStatePSS}{\mathbf{\bar v}}
\newcommand{\NextStateExPSS}{\mathbf{\hat v}}
\newcommand{\AcceptanceP}{r}
\newcommand{\RJMapping}{h_{ij}}

\newcommand{\Pinv}{S}
\newcommand{\Pfor}{S^{-1}}

\newcommand{\SceneJewelry}{\textsc{Jewelry}}
\newcommand{\SceneHorseRoom}{\textsc{Living Room}}
\newcommand{\SceneKitchen}{\textsc{Kitchen}}
\newcommand{\SceneStaircase}{\textsc{Staircase}}

\newcommand{\SceneBathroomTwo}{\textsc{Bathroom}}
\newcommand{\SceneBathroom}{\textsc{Salle de Bain}}
\newcommand{\SceneGlassOfWater}{\textsc{Glass of Water}}

\newcommand{\Paragraph}[1]{\textbf{#1}\ \ }

\newcommand{\Jac}[1]{J\big[#1\big]}

\newcommand{\misWeightPS}{w}

\hypersetup{draft}

\begin{document}
\title{Reversible Jump Metropolis Light Transport using Inverse Mappings}
\author{Benedikt Bitterli}
\affiliation{\institution{Dartmouth College}}
\affiliation{\institution{ETH Z{\"u}rich}}
\affiliation{\institution{Disney Research}}
\author{Wenzel Jakob}
\affiliation{\institution{\'Ecole Polytechnique F\'ed\'erale de Lausanne (EPFL)}}
\author{Jan Nov{\'a}k}
\affiliation{\institution{Disney Research}}
\author{Wojciech Jarosz}
\affiliation{\institution{Dartmouth College}}
\affiliation{\institution{Disney Research}}

\renewcommand{\shortauthors}{Bitterli et al.}


\begin{abstract}
We study Markov Chain Monte Carlo (MCMC) methods operating in primary sample space and their interactions with multiple sampling techniques. We observe that incorporating the sampling technique into the state of the Markov Chain, as done in Multiplexed Metropolis Light Transport (MMLT), impedes the ability of the chain to properly explore the path space, as transitions between sampling techniques lead to disruptive alterations of path samples. To address this issue, we reformulate Multiplexed MLT in the Reversible Jump MCMC framework (RJMCMC) and introduce \emph{inverse} sampling techniques that turn light paths into the random numbers that would produce them. This allows us to formulate a novel perturbation that can locally transition between sampling techniques without changing the geometry of the path, and we derive the correct acceptance probability using RJMCMC. We investigate how to generalize this concept to non-invertible sampling techniques commonly found in practice, and introduce \emph{probabilistic} inverses that extend our perturbation to cover most sampling methods found in light transport simulations. Our theory reconciles the inverses with RJMCMC yielding an unbiased algorithm, which we call \emph{Reversible Jump MLT} (RJMLT). We verify the correctness of our implementation in canonical and practical scenarios and demonstrate improved temporal coherence, decrease in structured artifacts, and faster convergence on a wide variety of scenes.
\end{abstract}

%
%

\ccsdesc[500]{Computing methodologies~Ray tracing}

%
%

\keywords{Ray tracing, photorealistic rendering}

%


\maketitle


\section{Introduction}
Monte Carlo rendering algorithms simulate the propagation of light by sampling random paths connecting the light source and a virtual sensor.
In scenes with complex materials, geometry, or lighting, the space of light paths is large and high-dimensional, but the subset of paths that contribute significantly to the image occupy only a narrow subspace. This makes rendering a notoriously difficult sampling problem even for state-of-the-art unbiased Monte Carlo techniques.

Markov Chain Monte Carlo (MCMC) rendering methods, such as \emph{Metropolis Light Transport} (MLT)~\citep{Veach:1997} generate a statistically dependent sequence of samples. This dependence augments the sampling process with a form of short-term memory that amortizes the search for important light paths by locally exploring newly discovered regions, which typically allows MCMC methods to handle such problematic scenarios more effectively. In its original path space formulation, MLT allows great flexibility in exploiting knowledge of the integration problem at hand, but requires crafting specialized perturbation strategies for different types of light transport (e.g.\ caustics, subsurface scattering). This increases its implementation complexity and reduces its generality in modern scenes that contain a combination of many effects.

\begin{figure*}[t]
    \centering%
    \includegraphics[width=\textwidth]{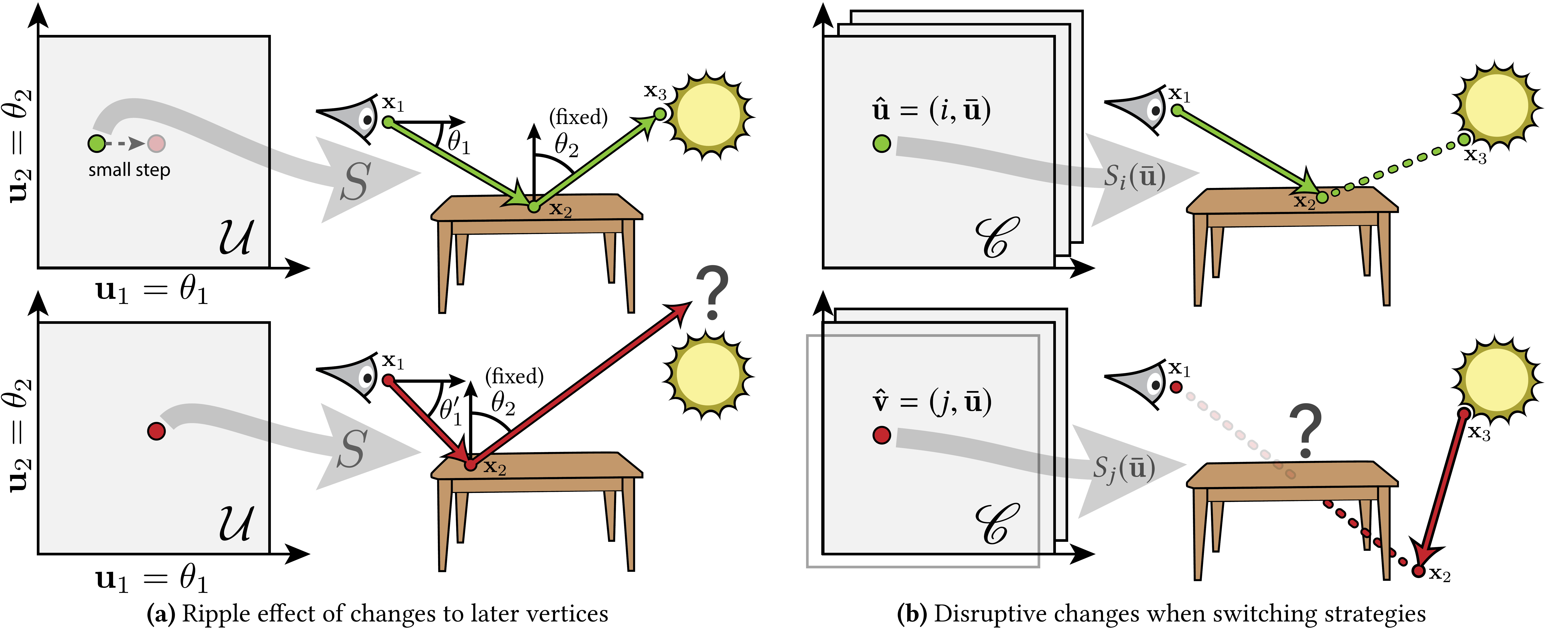}%
    \vspace*{-0.5\baselineskip}
    \caption{%
        \label{fig:mmlt-technique-change}%
        Fundamental issues of path sampling using primary sample space:
        \textbf{(a)} Perturbations in PSSMLT cause a ripple change that
        propagates to later vertices: here, a perturbation of the outgoing
        direction at the camera causes a large-scale change of the vertex on
        the light source $\mathbf{x}_3$. In such cases, it can be advantageous
        to switch to a different sampling strategy, for instance one that
        explicitly samples a position on a light source rather than
        intersecting it by chance. \textbf{(b)} Such strategy changes are
        possible using a multiplexed primary sample space such as that of MMLT.
        However, changing strategies generally leads to a large-scale change to
        the path geometry that causes the proposed path to be rejected with
        high probability.
        The RJMLT technique proposed in this paper introduces efficient
        strategy perturbations that leave the path geometry intact.
    }
\end{figure*}

\Citet{Kelemen:2002} later proposed a much simpler MCMC rendering approach---\emph{Primary Sample Space MLT} (PSSMLT)---which retrofits Metropolis sampling to existing Monte Carlo methods by treating them as abstract path samplers, and manipulating the random numbers they consume. This view brings many benefits, including ease of implementation and the ability to leverage existing sophisticated importance sampling strategies. \emph{Multiplexed MLT} (MMLT)~\citep{Hachisuka:2014} further improves efficiency by allowing the Markov Chain to adaptively select the bidirectional sampling techniques that have high contribution. Unfortunately, operating in the space of random numbers can also make these methods significantly less successful at locally exploring important regions of the state space. For instance, since most sampling schemes construct paths incrementally, a small change in the random numbers used by one vertex on the path generally leads to a ripple change that affects all subsequent vertices, as illustrated in  \figref{fig:mmlt-technique-change}(a). In addition, changing path sampling strategies (e.g.\ starting the path at the light vs.\ the camera) means that the random numbers for constructing a path are reinterpreted as input to a different sampling strategy, producing an entirely different path, as illustrated in \figref{fig:mmlt-technique-change}(b). Both scenarios turn small perturbations of the random numbers into large, disruptive changes to the path. This inhibits the ability of the Markov Chain to explore the state space locally and increases the likelihood of the chain getting ``stuck'' in small parts of the state space, exacerbating structured artifacts and temporal instability.

In this paper, we propose \emph{Reversible Jump MLT} (RJMLT), which partially bridges the gap between the flexibility of path-space MLT and the simplicity of PSSMLT-type methods. To accomplish this, we first recast (Section~\ref{sec:reversible-multiplexed-jumps}) MMLT in the framework of Reversible Jump MCMC~\citep{Green:1995}, which provides us a mathematical foundation for reasoning about the aforementioned problems. We then enrich the set of mappings available to PSSMLT methods with \emph{inverse} sampling techniques that turn light paths into the random numbers that would produce them. This allows us to formulate a new perturbation that can locally transition between sampling techniques for the same path (Section~\ref{sec:rmj}). We consider practical implications of such inverses (Sections~\ref{sec:inversion-problems}--\ref{sec:practical-inverses}) and reconcile them with the RJMCMC framework. We evaluate the correctness and efficiency of the resulting algorithm (Section~\ref{sec:results}), showing that we can achieve significantly higher acceptance rates than MMLT and reduce noise and the erratic convergence behavior of PSSMLT methods. Finally, we discuss  other ways that reversible jumps and inverse mappings can be used to improve PSSMLT methods in the future (Section~\ref{sec:conclusion}).

\section{Related Work \& Mathematical Background}
\label{sec:related+background}

We now discuss the most relevant prior work and establish a common mathematical notation to more firmly relate prior approaches to our contributions. We aggregate the most important terms of our notation in \tabref{tab:notation} for reference.

Current physically based rendering algorithms can all be expressed as approximations to a \emph{measurement equation}, which considers integrals of the form:
\begin{align}
    \Measurement_j = \int_\PathSpace \ReconFilter{j}(\xbar) \MContrib(\xbar) \, \dPath{\xbar} \punct{.} \label{eq:measurement-equation}
\end{align}
This equation computes the value of a measurement $\Measurement_j$ (usually, a pixel) in terms of an integral over all possible \emph{light paths}. The integrand is composed of the pixel reconstruction filter $\ReconFilter{j}(\xbar)$ and the contribution $\MContrib(\xbar)$ of light path $\xbar$ (note the bar notation, which indicates quantities organized as multiple vertices). The domain of all such light paths is \emph{path space} $\PathSpace$. We can decompose $\PathSpace$ further into subdomains $\PathSpace^k$ that contain only light paths of a fixed length $k$, which together form $\PathSpace = \bigcup_{k=2}^\infty \PathSpace^k$.

\begin{table}[t]%
\small
\vspace*{-1\baselineskip}
\caption{Table of notation}\label{tab:notation}%
\vspace*{-0.5\baselineskip}
    \begin{tabular}{ll}
    \toprule
        Symbol & Explanation \\
        \midrule
        $\PathSpace$ & Path Space \\
        $\PSSpace$ & Primary Sample Space \\
        $\PathSpace^k, \PSSpace^k$ & Spaces of paths of length $k$ \\
        $\PSSpaceSize_k$ & Size of $\PSSpace^k$ \\
        $\xbar\!=\!\xbarAt{1}\!\ldots\!\xbarAt{n}$, $\ybar\!=\!\ybarAt{1}\!\ldots\!\ybarAt{n}$ & Light paths\\
        $\ubar\!=\!\ubarAt{1}\!\ldots\!\ubarAt{n}$, $\vbar\!=\!\vbarAt{1}\!\ldots\!\vbarAt{n}$ & Random number vectors\\
        $\auxbar\!=\!\auxbarAt{1}\!\ldots\!\auxbarAt{n}$& Auxiliary variables for path inversion\\
        $\MContrib(\xbar)$ & Path contribution function \\
        $\pdf(\xbar)$ & PDF of sampling $\xbar$ \\
        $\MPSTarget(\xbar) = \MContrib(\xbar)/\pdf(\xbar)$ & PSSMLT importance function\\
        $\Pinv_i(\ubar)$ & $i$-th sampling technique of BDPT \\
        $\misWeightPSS_i(\ubar)$ & MIS weight of $i$-th technique \\
        $\pdfPSS_i(\ubar)$ & PDF of $i$-th technique \\
        $\MImportancePSS_i(\ubar)$ & $\MImportance(\Pinv_i(\ubar))$ \\
        $\MPSSTargetImportance_i(\ubar) = \misWeightPSS_i(\ubar) \MImportancePSS_i(\ubar)/\pdfPSS_i(\ubar)$ & MMLT importance function \\
        $\CurrentStateExPSS = (i, \ubar)$ & Multiplexed state with explicit technique\\
        $\ProposalDistribution(\ubar \rightarrow \vbar)$ & Proposal density of $\vbar$ from state $\ubar$ \\
        $\AcceptanceP(\ubar \rightarrow \vbar)$ & Probability of accepting proposal $\vbar$ \\
        $\Jac{A}$ & Jacobian Matrix of function $A$ \\
\bottomrule%
    \end{tabular}
\end{table}

\Paragraph{Monte Carlo Integration.}
Traditional Monte Carlo rendering---first pioneered with \citeauthor{Kajiya:1986}'s~\citeyear{Kajiya:1986} rendering equation and the path tracing algorithm---approximate
\equref{eq:measurement-equation} using the estimator:
\begin{align}
    \Measurement_j \approx \frac{1}{N} \sum\limits_{i=1}^N \frac{\ReconFilter{j}\big(\xbar^{(i)}\big) \MContrib\big(\xbar^{(i)}\big)}{\pdf\big(\xbar^{(i)}\big)} \label{eq:mc-estimator},
\end{align}
which averages $N$ independently sampled light paths $\xbar^{(i)}$, where $\pdf(\xbar^{(i)})$ is the probability density of the $i$-th path sample. These probability densities directly determine the variance of the estimator, and it is thus vital that $\pdf(\xbar^{(i)})$ is approximately proportional to the integrand to obtain an efficient integrator. Bidirectional variants of path tracing (BDPT)~\citep{Lafortune:1993,Lafortune:1996,Veach:1994} build on the reciprocal nature of light transport to construct partial light paths from both camera and light sources and connect them to yield large families of sampling strategies. For a length-$k$ path (i.e.\ with $k$ segments), BDPT considers $k+2$ different sampling techniques where each strategy importance samples different parts of the integrand. The resulting estimators are typically combined into a joint estimator using \emph{multiple importance sampling} (MIS)~\citep{Veach:1995}.



\Paragraph{Markov Chain Monte Carlo.}
MCMC Rendering algorithms such as MLT~\citep{Veach:1997} mark a significant departure from classical Monte Carlo rendering. By using local exploration, these methods can remarkably generate a correlated sequence of samples (light paths) with a density that is exactly proportional to an arbitrary non-negative function $\StationaryDistribution(\CurrentState)$.  Usually this is accomplished using Metropolis-Hastings~\citep{Hastings:1970, Veach:1997} steps: given current state $\CurrentState$, a \emph{proposal} state $\NextState$ is drawn from a proposal distribution $\ProposalDistribution(\CurrentState \rightarrow \NextState)$ which is accepted and becomes the next state of the Markov Chain with probability
\begin{align*}
    \AcceptanceP(\CurrentState \rightarrow \NextState) = \min \left\{1, \frac{\StationaryDistribution(\NextState) \ProposalDistribution(\NextState \rightarrow \CurrentState)}{\StationaryDistribution(\CurrentState) \ProposalDistribution(\CurrentState \rightarrow \NextState)}\right\}\!\punct{;}
\end{align*}
otherwise, it is rejected and the process repeats anew. Under relatively weak conditions on $\ProposalDistribution(\CurrentState \rightarrow \NextState)$, the distribution of Markov Chain states will converge to $\StationaryDistribution(\CurrentState)$.

\Paragraph{Path-Space MLT.}
In the original MLT algorithm by \citet{Veach:1997}, the Markov Chain operates on path space $\PathSpace$ and the target density $\StationaryDistribution$ is the path contribution $\MContrib(\xbar)$.
This method relies on a set of two mutation strategies~(which change a path significantly) and three perturbation strategies~(which change it only slightly). While mutations ensure that all states are reachable, perturbations are the main tool by which the Markov Chain locally explores the state space. Subsequent generalizations of these strategies extend the perturbations to arbitrary chains of specular interactions~\citep{Jakob:2013} and rough materials~\citep{Kaplanyan:2014,Hanika:2015}. Other variations of MLT include Energy Redistribution Path Tracing~\citep{Cline:2005}, which independently simulates a large number of Markov Chains that are stratified in image space, and Gradient-Domain MLT~\citep{Lehtinen:2013}, which estimates gradients and uses them to reconstruct the rendered image. A severe problem with MLT and variations thereof is that admissible light paths lie on a lower-dimensional subset of path space with non-Euclidean structure, which tremendously complicates the design of mutations and perturbations. Existing strategies only target specific physical effects, but modern scenes contain many interactions types that may even occur simultaneously on a single light path.

\Paragraph{Primary Sample Space MLT.}
PSSMLT~\citep{Kelemen:2002} takes a different approach that avoids these problems: it is based on the observation that each iteration of a classical Monte Carlo rendering algorithm (e.g.\ BDPT) consumes a limited number of random variates and uses them to sample one or more light paths. We can express this sampling process more formally in terms of an abstract sampling scheme $\Pinv(\ubar)$, which takes a vector $\ubar$ of random numbers drawn uniformly, and generates a light path. The domain of these vectors forms \emph{primary sample space} $\PSSpace$. Similar to path space, we can further decompose this space over path lengths, such that $\PSSpace = \bigcup_{k=1}^\infty \PSSpace^k$ and $\PSSpace^k = [0, 1]^{\PSSpaceSize_k}$.
The dimensionality $\PSSpaceSize_k$ is chosen to be large enough such that all light paths of length $k$ can be sampled by $\Pinv$. This allows us to rewrite \equref{eq:mc-estimator} explicitly in terms of the sampling scheme, using the relation $\xbarAt{i} = \Pinv\big(\ubarAt{i}\big)$.

PSSMLT operates the Markov Chain in primary sample space $\PSSpace$, instead of path space $\PathSpace$, and relies on the underlying rendering algorithm to convert such samples into light paths using the sampling scheme(s) at its disposal. The target function $\MPSSTarget(\ubar)$ then becomes the composition of the path contribution and sampling scheme,
\begin{align}
    \MPSSTarget(\ubar) = (\MImportance \circ \Pinv)(\ubar)/\pdfPSS(\ubar) \punct{.}\label{eq:mmlt-contribution}
\end{align}
Among its many advantages, PSSMLT admits simple symmetric perturbation strategies due to the Euclidean structure of primary sample space. Additionally, any reflectance model supported by the underlying rendering algorithm is automatically handled by the resulting MCMC algorithm. The generality of this approach has led to numerous applications, e.g.\ in the context of photon tracing with MCMC sampling~\citep{Hachisuka:2011}, vertex connection merging/unified path sampling~\citep{Sik:2016}, tempering techniques such as Replica Exchange~\citep{Kitaoka:2009}, methods that control the Markov Chain by simulating Hamiltonian dynamics~\citep{Li:2015}, and rendering algorithms steered by arbitrary importance functions~\citep{Hoberock:2010,Gruson:2016}. PSSMLT is most often combined with BDPT~\citep{Kelemen:2002}, which utilizes $k+2$ different sampling schemes $\Pinv_0(\ubar), \ldots, \Pinv_{k+1}(\ubar)$ to construct paths of length $k$ from shorter eye/camera subpaths. PSSMLT evaluates \equref{eq:mmlt-contribution} by generating two subpaths and determining the MIS-weighted contribution for all pairs of vertices. Unfortunately, this sequence of steps is fairly expensive, and many of the vertex pairs will generally contribute little to no energy.

\Paragraph{Multiplexed MLT.}
MMLT~\citep{Hachisuka:2014} addresses this problem by informing the Markov Chain about the availability of multiple sampling strategies. It does so by running separate Markov Chains for each path length $k$, in which the first dimension of primary sample space is used to select one of the $k+2$ techniques of BDPT. A perturbation in MMLT may therefore change both the vector of random numbers, as well as which sampling technique is used to map that vector into a light path. Each Markov Chain uses the modified target function
\begin{align}
    \MPSSTarget_j(\ubar) = w_j(\ubar) (\MImportance \circ \Pinv_j)(\ubar)/\pdfPSS_j(\ubar) \punct{,}
\end{align}
which incorporates the MIS weight $w_j(\ubar)$ of the technique. Thus, instead of evaluating all strategies all the time, this choice is adaptively made by the underlying Metropolis-Hastings iteration, allowing the method to focus computation on successful sampling techniques that carry significant energy from the light sources to the camera.

Despite these improvements, MMLT (like PSSMLT) still suffers from structured artifacts and temporal instability since strategy changes and small steps in primary sample space $\PSSpace$ lead to large, disruptive changes to the resulting light path (recall Figure~\ref{fig:mmlt-technique-change}), inhibiting the Markov Chain's ability to locally explore state space.

\Paragraph{Design space of MLT algorithms.}
At a high level, we consider PSSMLT and MLT to be extremes in a space involving a trade-off between the ``opaqueness'' of the underlying representation and the design complexity of the resulting algorithm: the random number streams of PSSMLT have no clear physical interpretation, but this also allows for PSSMLT's concise and general perturbation strategy. In contrast, MLT operates on path space, which is a direct encoding of the physical scattering process, but this comes at a cost of significantly increased complexity. MMLT resembles PSSMLT, though the special role of the first dimension of the random number stream makes the representation slightly less opaque. However, incorporating the sampling technique into the state of the Markov Chain inadvertently ties the random number vector to the particular technique for constructing the path and leads to a loss in flexibility, as switching to a different sampling technique effectively amounts to a large scale \emph{mutation} of the path (Figure~\ref{fig:mmlt-technique-change}b). Our proposed RJMLT technique lies squarely in between: by allowing an existing path to be turned back into the random numbers that produce it, the distinction between primary sample space and path space ceases to be significant. On the other hand, our method requires the availability of inverse functions for many components of the rendering system.

\Paragraph{Reversible Jumps.}
Our technique builds upon the Reversible Jump MCMC algorithm (RJMCMC)~\citep{Green:1995}, which was originally developed in the area of Bayesian Statistics. Bayesian applications of MCMC techniques involve sampling from posterior distributions that take some form of evidence (e.g. empirical observations) into account. RJMCMC addresses the case where \emph{multiple} candidate models could describe a given set of observations, and RJMCMC then extends the Metropolis-Hastings algorithm with mutations that transition between these different models. In our context, these models correspond to different bidirectional sampling strategies that can construct the same path using a different number of steps on the camera and light paths.



\section{Reversible Multiplexed Jumps}
\label{sec:reversible-multiplexed-jumps}

The efficiency gain of MMLT comes from its ability to spend less time on ineffective sampling strategies, but its choice of perturbation strategies inhibit the Markov Chain from transitioning between different sampling techniques. To better analyze this problem and potential solutions, we introduce a novel reformulation of Multiplexed MLT in the Reversible Jump MCMC~\citep{Green:1995} framework (RJMCMC). This allows us to reason about perturbations that change sampling techniques in terms of deterministic mappings between spaces. This same framework will later allow us to derive a novel perturbation that leverages inverses to reliably transition between sampling techniques.

We begin by explicitly separating the choice of sampling technique from the rest of primary sample space. The Markov Chain then operates in the product of a discrete and continuous space, $\ExtendedPSSpace^k = \{0, \ldots, k+1\} \times \PSSpace^k$ referred to as \emph{multiplexed primary sample space}. In the RJMCMC view, this space is decomposed into subspaces $\bigcup_{i=0}^{k+1} \ExtendedPSSpace_i^k$, where $\ExtendedPSSpace_i^k = \{i\} \times [0, 1]^{\PSSpaceSize_k}$ is the subspace of the $i$-th sampling technique for paths of length $k$.

The set of perturbations proposed in Multiplexed MLT will always perturb the position in primary sample space, and may additionally change the technique index. To simplify analysis, we will separate the two concerns and focus on a hypothetical \textsc{TechniquePerturbation} strategy, which keeps the position in primary sample space fixed and only attempts to transition to a different sampling technique. That is, given the current state $\CurrentStateExPSS = (i, \ubar)$ in $\ExtendedPSSpace^k_i$, the perturbation samples a technique index $j$ from the distribution $\ProposalDistribution(i \rightarrow j)$ and generates the proposal $\NextStateExPSS = (j, \ubar)$ in $\ExtendedPSSpace^k_j$. The distribution $\ProposalDistribution(i \rightarrow j)$ is left unspecified, but for simplicity we assume it to be symmetric, i.e.\ $\ProposalDistribution(i \rightarrow j) = \ProposalDistribution(j \rightarrow i)$.

In the RJMCMC view, we can model such a perturbation as a \emph{jump} between subspaces. The proposal is generated from the current state by a \emph{deterministic} mapping $\RJMapping: \PSSpace^k \rightarrow \PSSpace^k$ that relates points from one space to the other; that is, $\NextStateExPSS = (j,\,\RJMapping(\ubar))$. The RJMCMC acceptance probability for such a proposal is
\begin{align}
	\AcceptanceP(\CurrentStateExPSS \rightarrow \NextStateExPSS) = \frac{\MPSSTargetImportance_j(\RJMapping(\ubar))\ProposalDistribution(j \rightarrow i)}{\MPSSTargetImportance_i(\CurrentStatePSS)\ProposalDistribution(i \rightarrow j)} \left\vert\,\Jac{\RJMapping}(\CurrentStatePSS)\right\vert \punct{,} \label{eq:tperturb-acceptance}
\end{align}
where $\Jac{\RJMapping}$ is the Jacobian matrix of $\RJMapping$. For \textsc{TechniquePerturbation}, the mapping is trivial with $\RJMapping(\ubar) = \ubar$ and an identity Jacobian.
%
%

Since only the sampling technique is changed and not the random numbers, we would hope in the MMLT view that the acceptance probability will only depend on the ratio of MIS weights, i.e.\  on how well the proposed technique samples the current path compared to the current sampling technique. However, expanding and simplifying the terms in \equref{eq:tperturb-acceptance} yields
\begin{align}
    \AcceptanceP(\CurrentStateExPSS \rightarrow \NextStateExPSS) &= \frac{\MPSSTargetImportance_j(\CurrentStatePSS)}{\MPSSTargetImportance_i(\CurrentStatePSS)} \big\vert \,\mathbb{I}\, \big\vert\\
        &= \frac{\misWeightPS_j(\CurrentStatePSS)\,\MPSTargetImportance(\Pinv_j(\CurrentStatePSS))}{\misWeightPS_i(\CurrentStatePSS)\,\MPSTargetImportance(\Pinv_i(\CurrentStatePSS))}\punct{,} \label{eq:tech-perturb-ratio}
\end{align}
where $\mathbb{I}$ is the identity matrix.

\equref{eq:tech-perturb-ratio} exposes the main problem of this na\"ive perturbation: Even though the random number vector was not changed, the proposed state uses a \emph{different mapping} to transform that vector into a light path. In general, $\Pinv_i(\CurrentStatePSS) \neq \Pinv_j(\CurrentStatePSS)$, and it is likely that the proposed path (and therefore $\MPSTargetImportance(\Pinv_j(\CurrentStatePSS))$) differs from the current path by a significant amount (\figref{fig:mmlt-technique-change}). Such large changes are unlikely to be accepted, which impedes the Markov Chain's ability to transition between different sampling strategies. Multiplexed MLT additionally couples technique changes with perturbations of $\ubar$, but this does not address the problem observed in the na\"ive \textsc{TechniquePerturbation}: the algorithm suffers from low acceptance rates whenever a proposal changes strategies.

\section{Invertible Sampling Techniques}
\label{sec:rmj}

Ideally, a technique perturbation would leave the current path unchanged while switching techniques.
To do so, it must find a new point $\NextStatePSS$ in primary sample space such that $\Pinv_j(\vbar) = \Pinv_i(\CurrentStatePSS)$, and jump to $\NextStatePSS$ as part of the perturbation.

Assume for now that sampling techniques $\Pinv_i$ are well-behaved in the sense that they are \emph{diffeomorphisms}, i.e.\ all $\Pinv_i$ are smooth and possess a smooth inverse. Then, the desired perturbation is easily accomplished using the mapping $\RJMapping(\CurrentStatePSS) = \Pfor_j(\Pinv_i(\CurrentStatePSS))$. Here, we used $\Pfor_j(\xbar)$, which is an \emph{inverse sampling technique} that transforms a light path into the random numbers that would produce it. The Jacobian of this mapping is
\begin{align}
    \big\vert\Jac{\RJMapping}(\CurrentStatePSS) \big\vert &=
        \big\vert\,\Jac{\Pfor_j \circ \Pinv_i}(\CurrentStatePSS)\big\vert \\
     &= \big\vert\,\Jac{\Pfor_j}(\Pinv_i(\CurrentStatePSS))\big\vert \cdot \big\vert\,\Jac{\Pinv_i}(\CurrentStatePSS)\big\vert \\
     &= \big\vert\,\Jac{\Pinv_j}(\vbar) \big\vert^{-1} \cdot \big\vert \,\Jac{\Pinv_i}(\CurrentStatePSS)\big\vert \punct{,}
\end{align}
%
%
where the last step follows from the inverse function theorem. This allows us to write the Jacobian of the mapping in terms of the Jacobians of the sampling techniques\footnote{To side-step the subtlety that vertices of a light path $\xbar=\mathbf{x}_1\ldots\mathbf{x}_n\in\PathSpace$ reside on $2$D subspaces of $3$D, we assume that each vertex $\mathbf{x_i}$ can be \emph{locally} parametrized using an orthonormal tangent frame. The concatenation of these local parameterizations then yields a local parameterization of the neighborhood of a light path, which facilitates reasoning about the densities of sampling strategies. For instance, $\left|\Jac{\Pinv_i}(\CurrentStatePSS)\right|$ refers to the Jacobian determinant of the $i$-th sampling strategy, which captures the change in volume when a small volume element in $\PSSpace$ is mapped to $\PathSpace$. This is purely a theoretical concern so these local parameterizations are not required during implementation.}.

The Jacobian of a sampling technique is closely related to its PDF. Indeed, if $\Pinv_i(\CurrentStatePSS)$ is a diffeomorphism, then $\big\vert\Jac{\Pinv_i}(\CurrentStatePSS)\big\vert = \pdfPSS_i(\CurrentStatePSS)^{-1}$~\citep{Kelemen:2002}. The acceptance probability of such a proposal is
\begin{align}
    \AcceptanceP(\CurrentStateExPSS \rightarrow \NextStateExPSS) &= \frac{\ProposalDistribution(j \rightarrow i) \,\MPSSTargetImportance_j(\NextStatePSS)}{\ProposalDistribution(i \rightarrow j)\,\MPSSTargetImportance_i(\CurrentStatePSS)} \frac{\pdfPSS_j(\NextStatePSS)}{\pdfPSS_i(\CurrentStatePSS)}\\
    &= \frac{\ProposalDistribution(j \rightarrow i) \,\misWeightPSS_j(\NextStatePSS) \,\MContribPSS_j(\NextStatePSS) \,\pdfPSS_j(\NextStatePSS)^{-1}}{\ProposalDistribution(i \rightarrow j)\,\misWeightPSS_i(\CurrentStatePSS) \,\MContribPSS_i(\CurrentStatePSS) \,\pdfPSS_i(\CurrentStatePSS)^{-1}} \frac{\pdfPSS_j(\NextStatePSS)}{\pdfPSS_i(\CurrentStatePSS)} \\
    &= \frac{\ProposalDistribution(j \rightarrow i) \,\misWeightPSS_j(\NextStatePSS) \,\MContrib(\Pinv_j(\NextStatePSS))}{\ProposalDistribution(i \rightarrow j)\,\misWeightPSS_i(\CurrentStatePSS) \,\MContrib(\Pinv_i(\CurrentStatePSS))} \\
    &= \frac{\ProposalDistribution(j \rightarrow i) \,\misWeightPSS_j(\NextStatePSS)}{\ProposalDistribution(i \rightarrow j)\,\misWeightPSS_i(\CurrentStatePSS)} \punct{,} \label{eq:mrj-perturb-ratio}
\end{align}
where the last cancellation was obtained with $\Pinv_i(\CurrentStatePSS) = \Pinv_j(\NextStatePSS)$, which holds by construction.

\paragraph*{Discussion} \equref{eq:mrj-perturb-ratio} has several interesting properties. As desired, this acceptance ratio only depends on the relative MIS weight of the proposed and current technique. This allows the Markov Chain to easily transition to the optimal sampling techniques as it explores path space. Additionally, the path contribution function does not appear in the acceptance ratio, and expensive retracing of paths is not required. Notably, the remaining terms in the acceptance ratio can be derived from the current state \emph{before generating the proposal}. This allows us to select an \emph{optimal} proposal distribution with $\ProposalDistribution(i \rightarrow j) = \misWeightPSS_j(\CurrentStatePSS)$. This will cancel all remaining terms and achieve an acceptance ratio of 1.

We will refer to this perturbation as a \emph{reversible jump}, and the rendering method derived from it as \emph{Reversible Jump MLT}. The method proposed here can be summarized as:
\begin{enumerate}
    \item Choose a proposal technique $j$ with probability $\misWeight_j(\Pinv_i(\ubar))$
    \item Jump to proposal state $\NextStateExPSS = (j, \Pfor_j(\Pinv_i(\ubar)))$
    \item Always accept $\NextStateExPSS$
\end{enumerate}

\section{Non-Invertible Sampling Techniques}
\label{sec:inversion-problems}

The perturbation introduced in the previous section assumed that sampling techniques are smooth and invertible, but this does not always hold in practice. Sampling methods are not required to be injective and might therefore map several random number vectors to the same path, which introduces ambiguities when attempting to compute inverses. We focus on two cases that are relevant to light transport simulations:

\Paragraph{Ambiguous Intervals.}
Primary sample space methods generally apportion a constant number of dimensions to each vertex of a path, of which only a subset may be used at each bounce. Every unused dimension collapses an entire $[0, 1]$ interval in primary sample space onto the same path. Similarly, if a single variate $u_i$ is used to sample a discrete property (e.g. the index of the light source), an interval $u_i\in[a, b]$ may map to the same path. However, when turning an existing path into the random numbers that produce it, we must commit to a specific value of $u_i$.

\Paragraph{Mixtures of Sampling Techniques.}
The second problematic case involves sampling strategies with overlapping support---for instance, consider a simple diffuse-specular BRDF that samples either the diffuse or specular lobe according to a probability $\alpha_\mathrm{diffuse}$:
\begin{align}
    \Pinv_\mathrm{phong}(\mathbf{u}) = \begin{cases}
        \Pinv_\mathrm{diffuse}(u_2,\ldots),\qquad& \mathrm{if\ }u_1 < \alpha_{\mathrm{diffuse}}, \\
        \Pinv_\mathrm{specular}(u_2,\ldots),&\mathrm{otherwise.}
    \end{cases}
\end{align}
Following an interaction with the material, most scattered directions can be sampled using \emph{two} distinct vectors of uniform variates corresponding to interactions with the diffuse and specular lobes, respectively. Also, note how
the probability density of the sampling scheme is simply the average of the diffuse- and specular PDFs, $\pdf_\mathrm{phong}(\mathbf{u}) = (\pdf_\mathrm{diffuse}(\mathbf{u}) + \pdf_\mathrm{specular}(\mathbf{u}))\,/\,2$, while the Jacobian becomes discontinuous at $u_1=\alpha_\mathrm{diffuse}$:
\begin{align}
    \Jac{\Pinv_\mathrm{phong}}(\mathbf{u}) = \begin{cases}
        \Jac{\Pinv_\mathrm{diffuse}}(u_2,\ldots),\qquad&\mathrm{if\ } u_1 < \alpha_\mathrm{diffuse},\\
        \Jac{\Pinv_\mathrm{specular}}(u_2,\ldots),&\mathrm{otherwise.}
    \end{cases}
\end{align}
Mixtures of sampling techniques lead to non-diffeomorphic mappings, which break the relation between the Jacobian determinant and the probability density
$\big\vert\Jac{S}(\mathbf{u})\big\vert =
p(\mathbf{u})^{-1}$ that we used in the derivation of the acceptance probability for the invertible case in \equref{eq:mrj-perturb-ratio}.

%
%

\Paragraph{Extended Path Space.}
To address these issues, we will first introduce a conceptual construction that extends path space with auxiliary dimensions such that no information about the random numbers is lost during sampling, allowing paths to be inverted exactly. Afterwards, we will introduce probabilistic inverses as a practical solution to non-invertibility and reason about them within the RJMCMC framework.

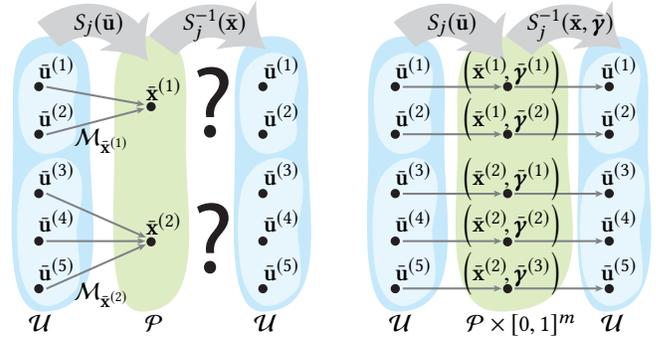
\begin{figure}[t]
\begin{overpic}[width=1\linewidth,unit=1mm]{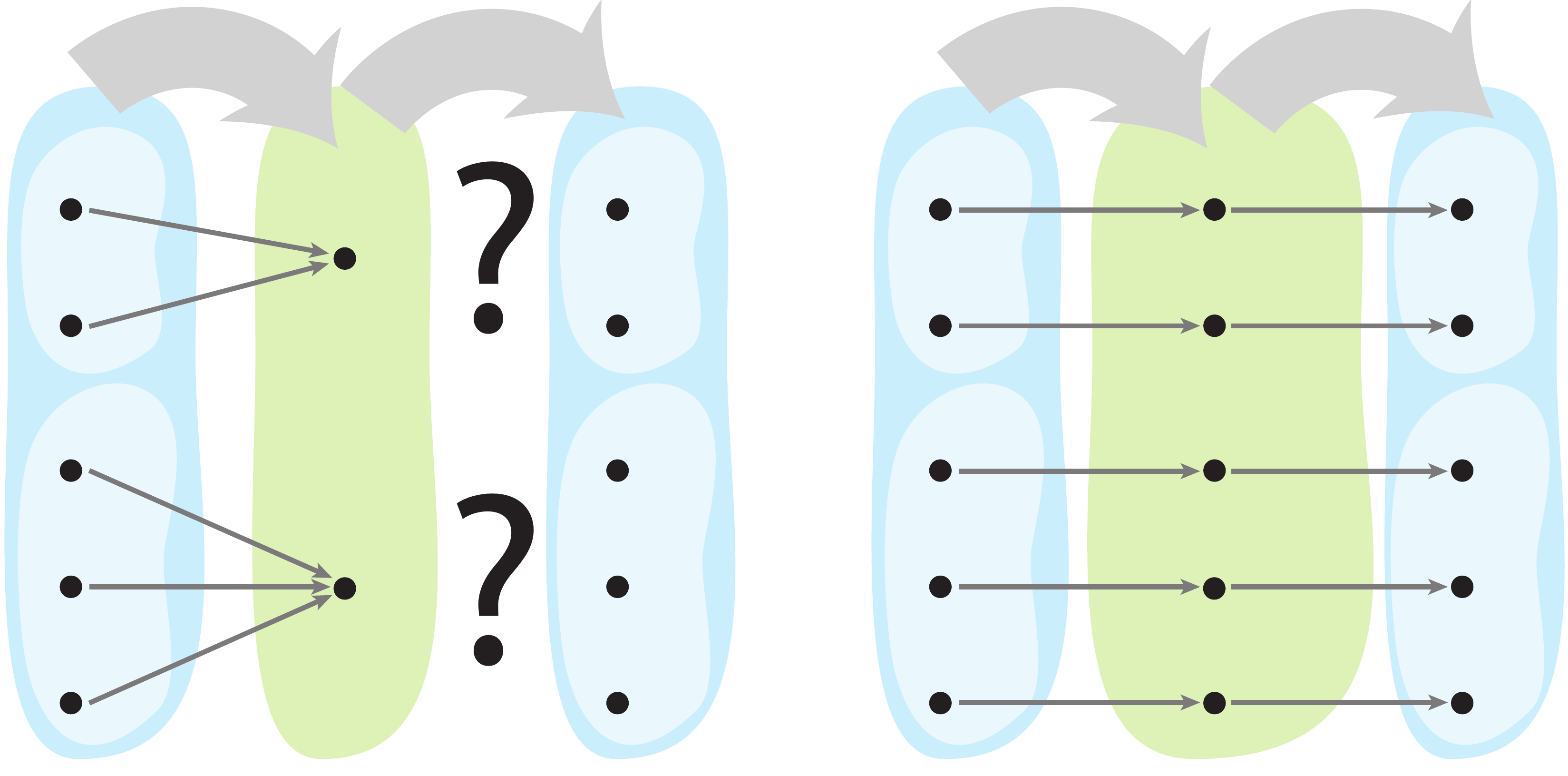}
    \put(4.5,37.3)  {$\ubar^{(1)}$}
    \put(4.5,30)    {$\ubar^{(2)}$}
    \put(4.5,20.8)  {$\ubar^{(3)}$}
    \put(4.5,13.5)  {$\ubar^{(4)}$}
    \put(4.5, 6)    {$\ubar^{(5)}$}
    \put(39.5,37.3) {$\ubar^{(1)}$}
    \put(39.5,30)   {$\ubar^{(2)}$}
    \put(39.5,20.8) {$\ubar^{(3)}$}
    \put(39.5,13.5) {$\ubar^{(4)}$}
    \put(39.5, 6)   {$\ubar^{(5)}$}
    \put(60, 37.3)  {$\ubar^{(1)}$}
    \put(60, 30)    {$\ubar^{(2)}$}
    \put(60, 20.8)  {$\ubar^{(3)}$}
    \put(60, 13.5)  {$\ubar^{(4)}$}
    \put(60,  6)    {$\ubar^{(5)}$}
    \put(92, 37.3)  {$\ubar^{(1)}$}
    \put(92, 30)    {$\ubar^{(2)}$}
    \put(92, 20.8)  {$\ubar^{(3)}$}
    \put(92, 13.5)  {$\ubar^{(4)}$}
    \put(92,  6)    {$\ubar^{(5)}$}
    \put(21.2,34)   {$\xbar^{(1)}$}
    \put(21.2,13)   {$\xbar^{(2)}$}
    \put(70.3,37.3) {$\left(\xbar^{(1)}\!,\auxbar^{(1)}\right)$}
    \put(70.3,30)   {$\left(\xbar^{(1)}\!,\auxbar^{(2)}\right)$}
    \put(70.3,20.8) {$\left(\xbar^{(2)}\!,\auxbar^{(1)}\right)$}
    \put(70.3,13.5) {$\left(\xbar^{(2)}\!,\auxbar^{(2)}\right)$}
    \put(70.3, 6)   {$\left(\xbar^{(2)}\!,\auxbar^{(3)}\right)$}
    \put(10, 27)    {$\mathcal{M}_{\xbar^{(1)}}$}
    \put(10, 3)     {$\mathcal{M}_{\xbar^{(2)}}$}
    \put(10,45)     {$\Pinv_j(\ubar)$}
    \put(27,45)     {$\Pfor_j(\xbar)$}
    \put(65,45)     {$\Pinv_j(\ubar)$}
    \put(80.5,45)   {$\Pfor_j(\xbar,\auxbar)$}
    \put(3, -2)       {$\PSSpace$}
    \put(21,-2)      {$\PathSpace$}
    \put(38,-2)      {$\PSSpace$}
    \put(59,-2)      {$\PSSpace$}
    \put(71,-2)      {$\PathSpace \times [0, 1]^m$}
    \put(92,-2)      {$\PSSpace$}
\end{overpic}
    \caption{Non-injective mappings $\Pinv_j(\ubar)$ (left) lead to ambiguities when transforming from path space to primary sample space. We sidestep this issue by ensuring all mappings are bijective (right); we extend the path space parameterizing its points by an additional parameter $\auxbar$ produced by $\Pinv_j(\ubar)$.}
    \label{fig:non-invert-sample-illustration}
\end{figure}

Suppose that the set of all random number vectors that map to a given path $\xbar$ is given by $\mathcal{M}_{\xbar} = \{\ubar\mid\Pinv_j(\ubar) = \xbar\}$.
In principle, we should be able to augment the inverse $\Pfor_j$ with an additional parameter $\auxbar \in [0, 1]^m$, that disambiguates the inverse so each pair of inputs $\Pfor_j(\xbar, \auxbar)$ maps to a single entry of $\mathcal{M}_{\xbar}$.
The role of $\auxbar$ will be to encode the ``extra''
dimensions in $\ubar$ that do not directly sample the path.
If such a map can be constructed (assuming a sufficiently large value of $m$), then we can perform reversible jumps with the following modifications:
sampling techniques generate points on an extended path space \mbox{$\PathSpace \times [0, 1]^m$}, and $\Pinv_i(\ubar)$ computes the pair $(\xbar, \auxbar)$, where $\xbar$ is the sampled path, and $\auxbar$ is the auxiliary input that would produce $\ubar$ when used in the inverse $\Pfor_i(\xbar, \auxbar)$; see~\figref{fig:non-invert-sample-illustration} for an illustration.

Using the extended path space, no information is lost when transitioning from or to primary sample space, enabling the use of RJMCMC to derive the acceptance probability
\begin{align}
    \AcceptanceP(\CurrentStateExPSS \rightarrow \NextStateExPSS) &= \frac{\MPSSTargetImportance_j(\vbar)\,\ProposalDistribution(j \rightarrow i)\ \big\vert \Jac{\Pinv_i}(\ubar)\big\vert}{\MPSSTargetImportance_i(\ubar)\,\ProposalDistribution(i \rightarrow j)\ \big\vert\Jac{\Pinv_j}(\vbar) \big\vert} \label{eqn:acceptance-rjmcmc-1}\\
    &= \frac{\MPSSTargetImportance_j(\vbar)\,\ProposalDistribution(j \rightarrow i)\ \left\vert \Jac{\Pfor_i}(\xbar, \auxbar)\right\vert^{-1}}{\MPSSTargetImportance_i(\ubar)\,\ProposalDistribution(i \rightarrow j)\ \left\vert\Jac{\Pfor_j}(\ybar, \auxbar) \right\vert^{-1}} \label{eqn:acceptance-rjmcmc-2}\\
	&= \frac{\MPSSTargetImportance_j(\vbar)\,\ProposalDistribution(j \rightarrow i)\ \left\vert\Jac{\Pfor_j}(\ybar, \auxbar) \right\vert }{\MPSSTargetImportance_i(\ubar)\,\ProposalDistribution(i \rightarrow j)\ \left\vert \Jac{\Pfor_i}(\xbar, \auxbar)\right\vert \label{eqn:acceptance-rjmcmc-3}} \punct{,}
\end{align}
where the step from \equref{eqn:acceptance-rjmcmc-1} to \eqref{eqn:acceptance-rjmcmc-2} follows from the inverse function theorem.
This construction provides a viable way of supporting non-invertible mappings, but it complicates the implementation as the auxiliary dimensions must be stored and propagated through the entire rendering system.

\Paragraph{Probabilistic inverses.}
We rely on a much lighter-weight solution to resolve ambiguities during path sampling: whenever multiple inverses are available, we simply randomly select one. This is realized by combining strategy change perturbations with an additional step that samples $\auxbar\in[0,1]^m$ from a uniform distribution. This modification to the proposal results in extra unit factors in the acceptance probability that cancel, hence the acceptance probability of a probabilistic inverse is still given by Equation~\eqref{eqn:acceptance-rjmcmc-3}. Since the auxiliary vector $\auxbar$ is re-sampled as part of every strategy perturbation, it is no longer part of the Markov Chain's state.

\section{Inverses in Practice}
\label{sec:practical-inverses}
In the previous section, we derived a general framework for handling path sampling methods that are not strictly invertible, and reconciled these ideas with the RJMCMC framework. In this section, we will now discuss inverses in more concrete terms and describe a few simple building blocks that allow correct inversion of many sampling methods used in light transport simulations.
We will also discuss how to compute the required Jacobians in more detail.

In practice, light paths are almost always sampled incrementally in a random walk. This corresponds to chaining a series of $\stCount$ low-dimensional sampling techniques, and we can write
\begin{align}
    \Pinv_i(\ubar) = \left(g_1\left(\ubarAt{1}\right), g_2\left(\xbarAt{1}, \ubarAt{2}\right), \ldots, g_\stCount\left(\xbarAt{\stCount-1}, \ubarAt{\stCount}\right)\right) \punct{,}
\end{align}
where $\xbar_l=\xbarAt{1}\ldots\xbarAt{l}$ is the path up to vertex $\xbarAt{l}$, and $g_l(\xbarAt{l-1}, \ubarAt{l})$ is the $l$-th sampling technique along the path, using a subset of the random numbers in $\ubar$. Inverting $\Pinv_i$ then reduces to inverting the individual techniques $g_l$, in an analogous ``inverse random walk''. The Jacobian determinant required by Equation~\eqref{eqn:acceptance-rjmcmc-3} turns into a product of determinants for each step, i.e.,
\begin{align*}
    \left\vert\,\Jac{\Pfor_i}(\xbar, \auxbar)\right\vert =
    \prod_{l=1}^q
    \left\vert\,\Jac{g_l^{-1}}\left(\xbarAt{l}, \auxbarAt{l}\right)\right\vert
    \punct{.}
\end{align*}
In order to maintain an optimal acceptance ratio and for \equref{eq:mrj-perturb-ratio} to hold, we wish for the individual Jacobians $\vert J[g_l^{-1}](\xbarAt{l}, \auxbarAt{l})\vert$ to be equal to the PDF of the corresponding mapping $g_l$. In the following, we will focus on a single sampling technique $g$, and drop the subscripts for ease of notation.

\Paragraph{Inversion method.}
Mappings based on the \emph{inversion method} form the basic
building blocks of many sampling techniques and are easily handled. Such
techniques draw samples from a probability distribution $p$ by mapping uniform
variates through the inverse CDF $P^{-1}$. These mappings are invertible by construction, and to support them in our system we simply require an additional implementation of the CDF $P$, which serves as a (non-probabilistic) inverse. The Jacobian determinant
$\left\vert\,\Jac{g^{-1}}\left(\mathbf{x}, \bm{\gamma}\right)\right\vert$ of this inverse is  equal to the PDF $p$ of the sampling technique.

In the remainder of this section, we turn to the two sources of non-invertibility
discussed in \secref{sec:inversion-problems}.


\subsection{Ambiguous Intervals}
\label{sec:ambiguous-intervals}
Ambiguous interval arose whenever a dimension $u_i$ in primary sample space could take on any value on an interval $[a, b]$ without changing the generated path.
Constructing a probabilistic inverse for this case is fortunately easy:
we generate a uniform variate $\auxScalar\in[0,1]$ and set $u_i=g^{-1}(\auxScalar)$, where
\begin{align}
    g^{-1}(\auxScalar) = a + \auxScalar \cdot (b - a) \quad\text{and}\quad \Jac{g^{-1}}(\auxScalar) = b - a \punct{.}\ \label{eq:ambiguous-intervals}
\end{align}
For entirely unused dimensions of primary sample space this reduces to $g^{-1}(\auxScalar) = \auxScalar$ with a unit Jacobian determinant; that is, unused dimensions are simply uniformly re-sampled during inversion and do not influence the acceptance probability.

\subsection{Mixtures of Sampling Techniques}
\label{sec:sampling-mixtures}


Suppose that $g$ consists of a combination of sampling techniques $g_1, \ldots, g_n$ selected at random, where technique $g_i$ is chosen with probability $\alpha_i$. We assume
that the technique index $t$ is chosen by the primary sample $u_1$ such that $\alpha_1 + \ldots + \alpha_{t-1} \leq u_1 < \alpha_1 + \ldots + \alpha_t$.

We now propose a probabilistic inverse that resembles this sampling procedure. First, we randomly select a technique index $t$ from a (yet undisclosed) discrete distribution $\ProposalDistribution(t)$. We then invert the sample assuming that it was generated by the $t$-th technique. This disambiguates both which interval the variate $u_1$ falls into, as well as which mapping $g_t^{-1}$ should be used. For a fixed $t$, the resulting inverse and Jacobian determinant are then
\begin{align}
    g^{-1}(\mathbf{x}, \aux) &= \big(\alpha_1 + \ldots + \alpha_{t-1} + \gamma_{1} \cdot \alpha_t, \;\; g_t^{-1}(\mathbf{x}, \gamma_{2}, \ldots)\big) \punct{,} \label{eq:layered-material-inverse}\\
    \big\vert\Jac{g^{-1}}(\xbar, \auxbar)\big\vert &= \alpha_t \cdot \big\vert\Jac{g_t^{-1}}(\xbar, \auxbarAt{2} \ldots)\big\vert\punct{.} \label{eq:layered-material-jacobian}
\end{align}
%
%
\begin{figure*}[t]
    \hspace*{-2mm}\input{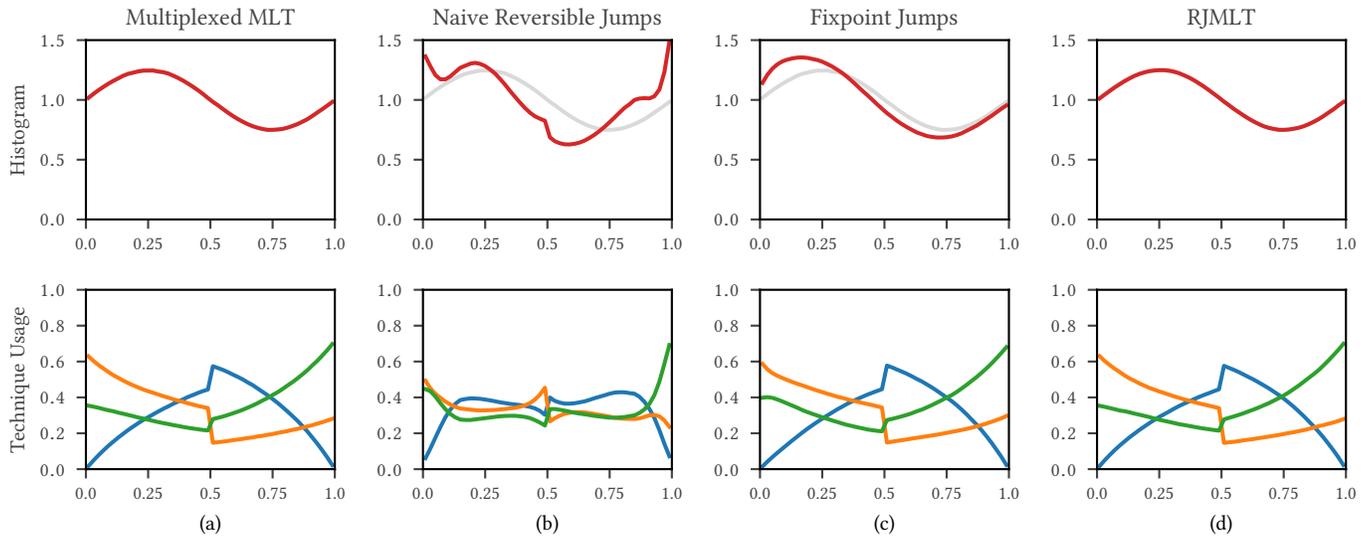}\\[-5mm]
    \caption{In a simplified 1D scenario, we compute the stationary distribution (top row) and average use of three available sampling techniques (bottom row) for four different Markov Chain integrators. We demonstrate the correctness of our full approach (d) compared to the Multiplexed MLT ground truth (a). Neglecting the necessary Jacobians (b) or improperly parametrizing inverses (c) leads to skewed distributions, demonstrating the importance of our full theory.}\label{fig:1d-results}
\end{figure*}

This presence of a technique index $t$ extends the proposal state generated by RJMLT: In addition to selecting the $j$-th technique of BDPT as MMLT does, the strategy perturbation also selects which of the $n$ sampling techniques should be used to invert $g$, which
yields a slightly modified acceptance probability
\begin{align}
    \AcceptanceP((\CurrentStateExPSS, t_{\ubar}) \rightarrow (\NextStateExPSS, t_{\vbar})) =
    \AcceptanceP(\CurrentStateExPSS \rightarrow \NextStateExPSS) \frac{T(t_{\ubar})}{T(t_{\vbar})}.
    \label{eq:mixture-mapping-acceptance}
\end{align}
The last step is to pick a concrete distribution $T$. Any distribution that samples
strategy $t$ with nonzero probability if it could potentially have produced
$\mathbf{x}$ is in principle admissible. 

We use
\begin{align}
    \ProposalDistribution(t) = \frac{\alpha_t \cdot \big\vert\Jac{g_t^{-1}}(\xbar, \auxbar)\big\vert}{\sum_{s=1}^n \alpha_s \cdot \big\vert\Jac{g_s^{-1}}(\xbar, \auxbar)\big\vert} \punct{,} \label{eq:mixture-discrete-distribution}
\end{align}
which cancels out the Jacobian \eqref{eq:layered-material-jacobian}
in the acceptance ratio, and results in an acceptance rate of 1.\footnote{$T$ has an intuitive interpretation: If $g_t$ is itself not a nested mixture of techniques, then $T(t)$ is simply the discrete probability that $\xbar$ was generated by the $t$-th technique.}

\Paragraph{Final algorithm}
\label{sec:jacobian-computation}
For completeness, we now state the outline of our complete strategy perturbation:
\begin{enumerate}
    \item Choose a proposal technique $j$ with probability $\misWeight_j(\Pinv_i(\ubar))$
    \item Jump to proposal state $\NextStateExPSS = (j, \Pfor_j(\Pinv_i(\ubar)))$
        \begin{itemize}
        \item If there are ambiguous intervals for elements of $\vbar$, sample them uniformly (Equation~\ref{eq:ambiguous-intervals})
        \item If $\Pinv_j$ uses a mixture of sampling techniques, select one randomly according to $T$ (Equation ~\ref{eq:mixture-discrete-distribution})
        \end{itemize}
    \item Always accept $\NextStateExPSS$.
\end{enumerate}
There is one small caveat to step (3): in rare cases, it might be impossible to invert a path due to numerical error. We detect such cases by checking if the path cannot be sampled by the method that we wish to invert (for example, if a direction lies in the wrong hemisphere), and reject proposals in this case.

\section{Results}
\label{sec:results}

We validate our theory with two implementations of our proposed method. Our first implementation performs MCMC integration in a simplified 1D scenario. The purpose of this simulation is not to compare performance across different techniques, but to demonstrate the correctness of our approach. We find that the high dimensionality of the light transport problem and the use of large steps can often mask subtle biasing issues of Markov Chain rendering methods, whereas these are immediately apparent in the 1D case.

Our second implementation adds reversible jumps to a general ray tracing renderer as an auxiliary perturbation on top of Multiplexed MLT. We evaluate the performance of this implementation in detail in \secref{sec:tungsten-results}.

\subsection{1D Markov Chain Integrator}

Our simulation performs Markov Chain integration on the $[0, 1]$ interval with a sinusoid target distribution (\figref{fig:1d-results} (a), top). Each integration method has three sampling techniques at its disposal with PDFs corresponding to: a triangular distribution (a), a step function distribution (b), and a mixture distribution (c) of a sinusoid and a linear function, as illustrated below.
\begin{figure}[h]
\hspace*{-2mm}%
\input{figures/1d-plots/Techniques.pgf}\\[-5mm]
\end{figure}

The primary sample space in this scenario has three dimensions: One used for selecting the currently used technique, one for sampling a position and one for selecting a subtechnique in mixture distributions (only utilized in the third sampling technique). We demonstrate the results for this setup in \figref{fig:1d-results}.

The histogram of the Markov Chain states (top row) should converge to the sinusoid target distribution if the algorithm is correct. We also record a secondary histogram that keeps track of which sampling techniques were used by the Markov Chain on average at particular points in space (bottom row). For the integration methods considered here, this histogram should converge to the MIS weights of the sampling techniques.

We implemented the equivalent of Multiplexed MLT in this 1D framework as a baseline integrator (\figref{fig:1d-results} (a)), which represents the ground truth for both histograms. In order to amplify potential issues in the perturbations, we disable large steps for all integrators, and only allow the Markov Chain to explore the 1D space using small steps and reversible jumps.

To demonstrate the importance of correctly performing reversible jumps, we implement two versions of our algorithm that use only parts of our theory. The first integrator performs reversible jumps without taking into the account the Jacobian when transitioning between subspaces, and omits the right-most determinant term in (Eq.\ \eqref{eqn:acceptance-rjmcmc-3}). Unlike traditional perturbations, reversible jumps are fully deterministic aside from selecting one of the available inverses, and one may be tempted to treat them as a discrete transition with trivial acceptance probability. However, neglecting to incorporate the Jacobian of the transition leads to a distorted stationary distribution (\figref{fig:1d-results} (b)) and a skewed use of the different sampling techniques.

The second integrator includes Jacobians in the acceptance probability (Eq.\ \eqref{eqn:acceptance-rjmcmc-3}) and inverts sampling mixtures nearly correctly (Eq.\ \eqref{eq:layered-material-inverse}), but does not fully parametrize $\mathcal{M}_{\xbar}$. Because the third dimension of primary sample space is used to select a subtechnique, an ambiguous interval arises when inverting the mixture distribution. Rather than parametrizing these intervals as described in \secref{sec:ambiguous-intervals}, this integrator always returns a fixed point inside the interval. Although this approach may not immediately appear incorrect, computing inverses in this manner leads to a biased distribution (\figref{fig:1d-results} (c)).

Finally, we implemented our full approach, using both Jacobians and parametrized inverses. Incorporating our full theory leads to a correct stationary distribution, as shown in \figref{fig:1d-results} (d).

\subsection{General Ray Tracing Renderer}\label{sec:tungsten-results}

Our most general implementation adds RJMLT to an existing rendering system. To ensure reproducibility, we will release our implementation as open source following publication. We implement reversible jumps as an additional perturbation on top of Multiplexed MLT. At each step, the Markov Chain selects either a large step, small step or a technique change with a fixed probability distribution (10\%, 85\% and 5\%, respectively). Unlike Multiplexed MLT, we explicitly separate the technique index from the rest of primary sample space, such that small step perturbations may not change the sampling technique.

We tested our implementation on an array of 8 different test scenes, and gather a series of metrics to evaluate our approach. Because MLT-based rendering methods can only compute a correct solution up to a global scaling factor, we run a path tracer on each scene for several hours to compute noise-free estimates of these factors and supply them to both MMLT and RJMLT. This allows us to exclude differences in scaling factors from the comparison.

\begin{figure*}[t!]
    \hspace*{-4mm}\includegraphics{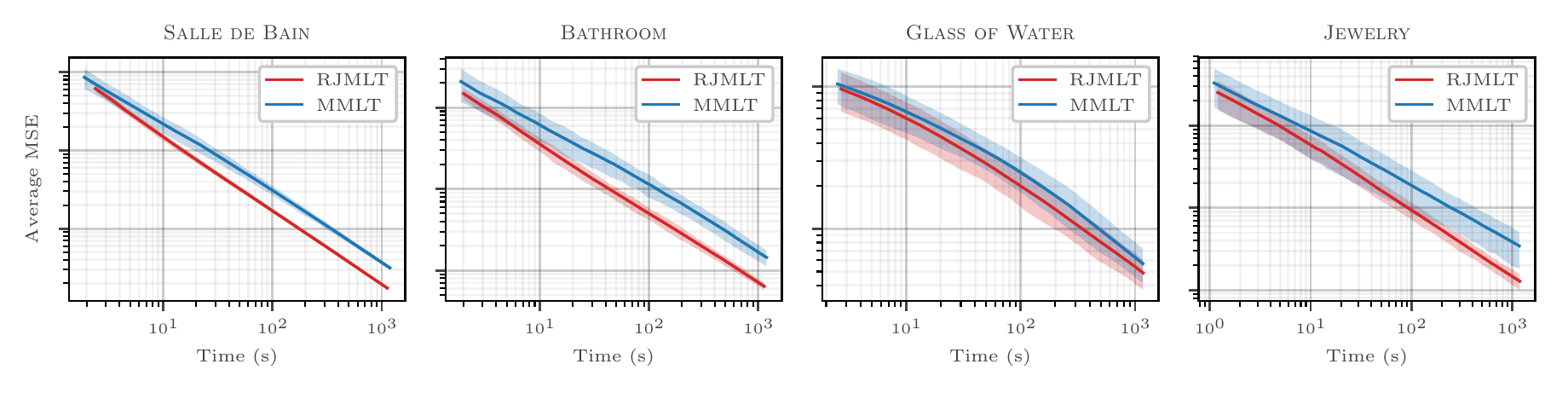}\\[-3mm]
    \hspace*{-4mm}\includegraphics{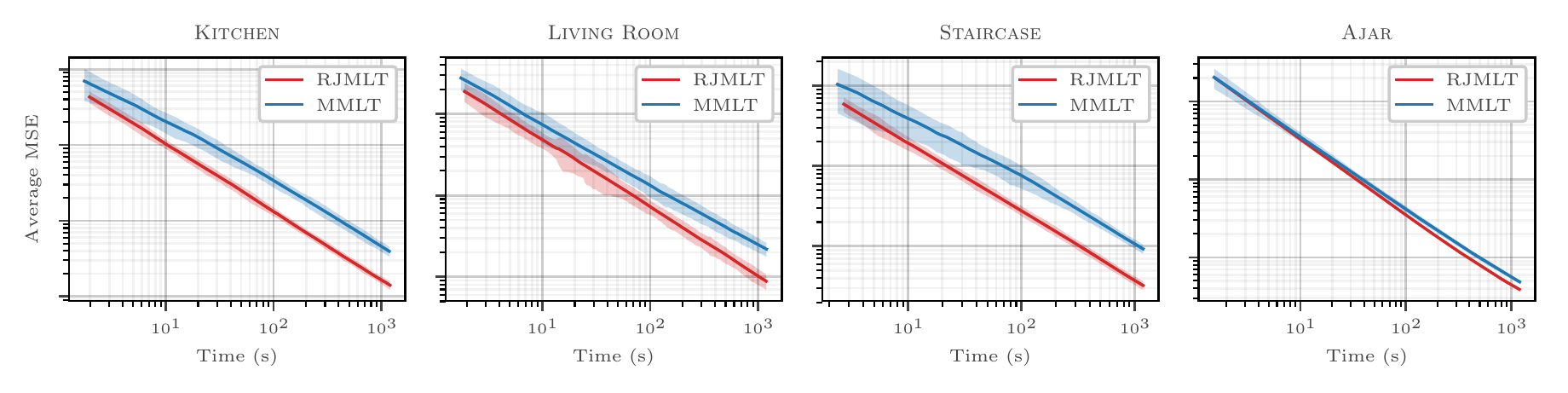}\\[-5mm]
    \caption{To compare the convergence of our method to previous work, we measure the MSE over time of 50 independent runs of RJMLT and MMLT in all 8 of our test scenes. We visualize the average MSE curve over time (thin curve), as well as the standard deviation of the MSE across all 50 individual runs (shaded areas). At equal time, RJMLT both has a lower MSE on average, as well as less variation across different runs.}\label{fig:mse-plots}
\end{figure*}

The use of correlated samples can make a fair comparison between MCMC-based rendering methods challenging. Because Markov Chains generate most samples with local exploration, integration error can manifest in different forms, such as ``splotches'', streaks, wrongly positioned caustics or entirely missing transport modes. The presence of many such artifacts is symptomatic of the Markov Chain ``getting stuck'' in local maxima, and suggests that inappropriate perturbations are employed by the integrator. These artifacts have implications for both the visual appearance and quantitative convergence, and we provide a series of comparisons to highlight the differences between RJMLT and MMLT:

\Paragraph{Convergence Plots.} Inappropriate perturbations lead to a slower convergence of the Markov Chain to its target distribution. To perform a quantitative analysis of this fact, we track the MSE (compared to a path traced reference image) of both RJMLT and MMLT as a function of render time. Because of the use of correlated samples, a single run of an MCMC integrator may not be representative, and we run 50 instances of each integrator with different random seeds to get a measure of its average behavior. In addition, the \emph{variation} of the MSE between runs is an indication of the stability of the integrator. We plot graphs of both of these statistics in \figref{fig:mse-plots}. The thin line represents the average MSE over time, whereas the shaded area visualizes the standard deviation of the MSE across \emph{all} 50 runs. In all test scenes, RJMLT has a lower average MSE than MMLT at equal render time, sometimes significantly. Additionally, RJMLT exhibits less variation between runs than MMLT, which suggests that our proposed perturbation is more effective at exploring potential light paths than MMLT small steps.

\begin{figure}[t!]
    \vspace*{-1\baselineskip}
    \hspace*{-1mm}\input{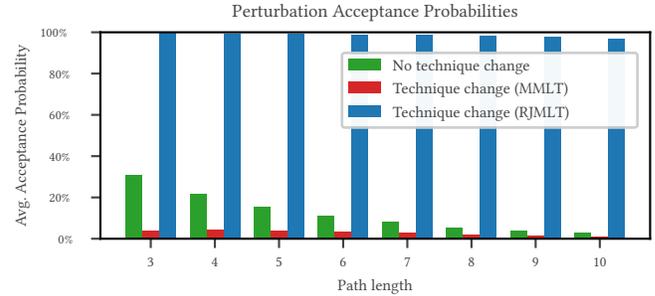}\\[-5mm]
    \caption{We visualize the average acceptance probability of perturbations in the \SceneHorseRoom~scene, broken down over path length. To demonstrate the benefit of our method, we differentiate between proposals that change the sampling technique, and proposals that leave it unchanged. For Multiplexed MLT, proposals that attempt to change the sampling technique have a significantly lower acceptance probability compared to proposals that do not, and it is difficult for the Markov Chain to transition between sampling techniques. Conversely, technique changes in Reversible Jump MLT are nearly always accepted.}\label{fig:acceptance-ratio}
    \vspace*{-1\baselineskip}
\end{figure}

\Paragraph{Acceptance Probabilities.} A direct measure of the performance of a perturbation is the rate at which its proposals are accepted. We track the average proposal acceptance rates in the \SceneHorseRoom~scene, and compare the results in \figref{fig:acceptance-ratio}, broken down over path length. We differentiate between three different perturbations: MMLT small steps that leave the sampling technique unchanged, MMLT small steps that change the technique, and our proposed RJMLT perturbations. RJMLT small steps and MMLT small steps without technique change have identical behavior, and we only show the MMLT results. As the path length increases, the acceptance rates of small step proposals decreases rapidly. Additionally, MMLT proposals that change the technique index have a significantly lower acceptance probability, dropping as low as 1\% at path length 10. In contrast, our proposed perturbations achieve average acceptance probabilities close to 100\% across all path lengths. The slight drop for longer paths stems from perturbations that could not perfectly invert a path due to floating point inaccuracy, and were rejected by our algorithm.

\begin{figure*}[t!]
    \definecolor{gainsboro}{rgb}{0.89, 0.0, 0.13}
    \definecolor{charcoal}{rgb}{0.93, 0.8, 0.0}
    \definecolor{arrowcolor}{rgb}{1, 1, 1}
    \linethickness{0.3mm}
    \centering
    \renewcommand{\tabcolsep}{0pt}
    \renewcommand{\arraystretch}{0}
    \input{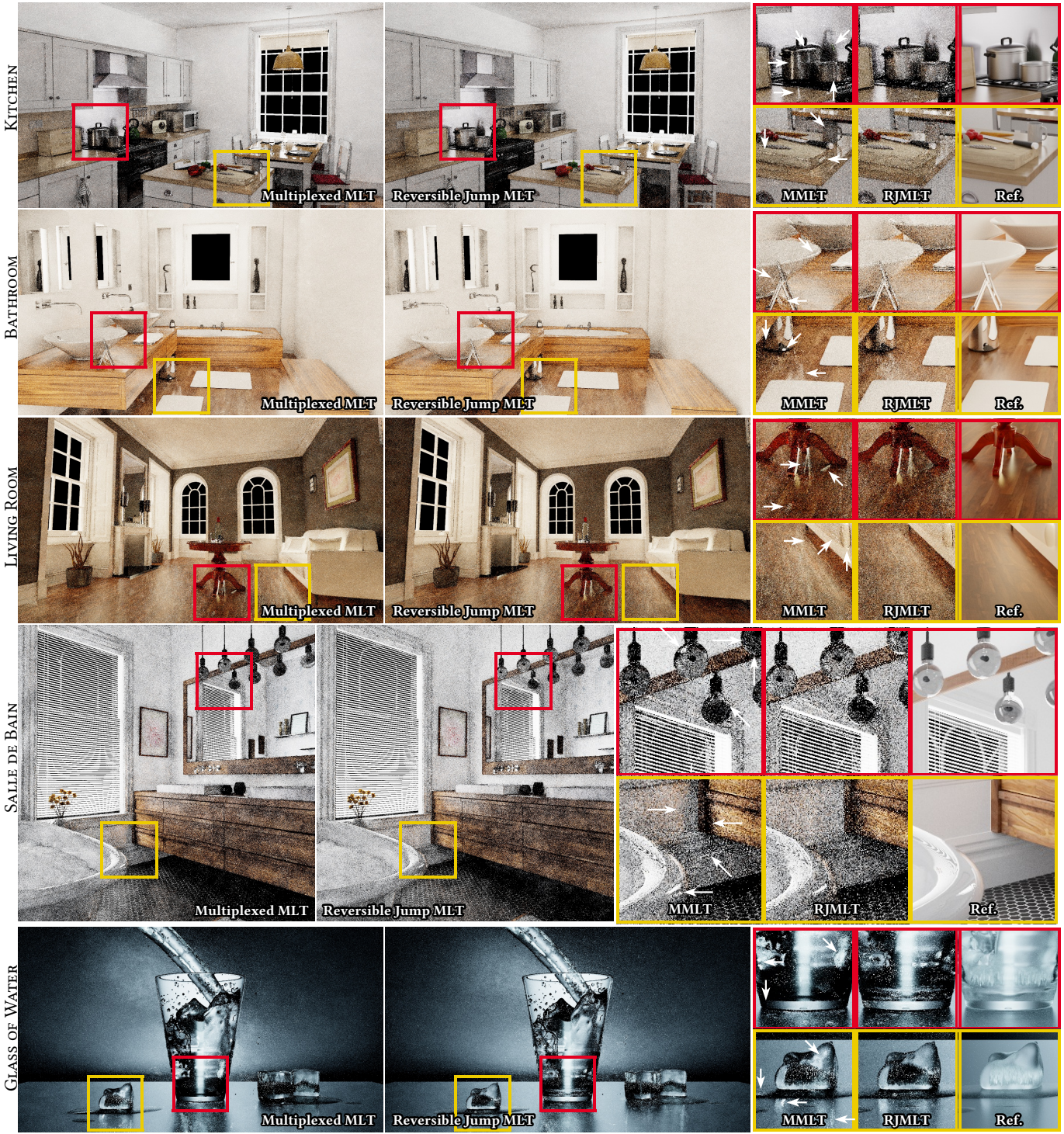}
    \caption{We compare Multiplexed MLT (left) to Reversible Jump MLT (middle) across 5 different scenes at equal render time. The scenes feature complex illumination and occlusion, glossy caustics and long light path lengths. Because of its ability to transition between sampling techniques easily, RJMLT has fewer artifacts (streaks, splotches, wrongly positioned caustics) than Multiplexed MLT at equal render time. Such artifacts are symptomatic of the Markov Chain ``getting stuck'' and exploring a small part of path space for too long.\label{fig:results}}
\end{figure*}

\Paragraph{Temporal Stability.} A drawback of correlated sampling is that noise is spatially coherent. This may not be immediately apparent in still images, but can cause visually disturbing flickering artifacts in rendered animations. To provide a visual comparison of the temporal behavior of MMLT and RJMLT, we render 50 images for each scene and integrator with different random seeds. We compose the renderings into a video, and provide a web-based comparison tool in our supplemental material to compare the temporal videos side-by-side. In all scenes, RJMLT exhibits more stable and visually pleasing noise behavior across runs than MMLT, and in some scenes (e.g.\ \SceneStaircase, \SceneJewelry, \SceneBathroom, \SceneKitchen) significantly so.

\Paragraph{Equal-time renderings.} Finally, we also provide renderings of RJMLT and MMLT after 5 minutes of render time in \figref{fig:results}. We show a subset of our scenes, and provide two insets for each scene highlighting interesting features. Because of the structured nature of MLT noise, it is difficult to judge renderings of the methods by their noise level alone. We have added arrows to the MMLT insets pointing out areas of significant error. These include e.g.\ streaks on the pots and cutting board in the \SceneKitchen~scene; streaky reflections and hard-edge shadows in \SceneHorseRoom; wrongly positioned or missing caustics in \SceneBathroom; and significant splotches in \SceneGlassOfWater. We include full-size renderings of all our methods in the supplemental material, along with a web-based image comparison tool. We also include a heatmap image of the MSE in each scene, averaged over 50 runs of the integrators.

\section{Conclusion}
\label{sec:conclusion}

We introduced Reversible Jump MCMC to the field of light transport, and presented a reformulation of Multiplexed MLT in this framework. Our analysis showed that technique changes in MMLT amount to disruptive changes to the light path, and predicted difficulties in transitioning between sampling techniques. We introduced the concept of inverse sample functions, which transform light paths into the random numbers that would produce them. Using these inverses, we constructed a novel perturbation, which we call \emph{Reversible Jump MLT}, that allows the Markov Chain to easily transition between sampling techniques. We perform an in-depth mathematical analysis of this perturbation, and derive the correct acceptance probability using the RJMCMC framework. We describe how to construct probabilistic inverses, and reconcile them with the RJMCMC framework. Two implementations confirm the correctness of our method, and detailed metrics show improved temporal coherence, faster convergence and decrease in artifacts of our method compared to MMLT.

\subsection{Limitations}

Even though path sampling methods are not always invertible, we described general mathematical tools for handling these issues and how to express these ideas in the RJMCMC framework. We found the concrete techniques we described sufficient to handle nearly every sampling scheme we encountered in practice. However, there are a few notable exceptions that may prevent inversion for certain kinds of paths, and our perturbation will fail in such cases. This does not affect the correctness of the method, but it may impact convergence if these paths dominate transport in the scene.

Rejection sampling is an alternative recipe for deriving sampling algorithms for general target distributions. Sampling methods derived in this manner, such as Woodcock tracking~\citeyear{Woodcock:1965}, consume a potentially unbounded set of random numbers and are virtually impossible to invert. However, the gains obtained from applying PSSMLT based methods on top of such a sampling scheme are questionable, and we believe this not to be a practical issue.

Certain specialized sampling schemes, such as the Box-Muller transform~\citeyear{Box:1958}, produce more than one sample for one set of inputs. Surplus samples are usually cached and reused in future sample queries, introducing a dependence between paths. Given only a single sample, it is not clear how to invert such a scheme and how the correct acceptance ratio should be computed. Our current implementation will simply reject perturbations involving such schemes.

\subsection{Future Work}

There are a number of interesting directions in which our work could be extended. A new class of perturbations is enabled by the ability to transition between path space and primary sample space, of which RJMLT is only one possibility. For example, the existence of inverses could allow path space methods to temporarily transition to primary sample space, perform a perturbation and transition back to path space. This reintroduces some of the benefits of PSSMLT, such as a better local parametrization of paths, to path space methods. Conversely, inverses could enable crafting path-space-style perturbations and mutations in a PSSMLT framework.

In addition to the theory presented here, Reversible Jump MCMC also provides a framework to reason about \emph{dimension jumps}, i.e.\ transitions between subspaces of different dimensions akin to the bidirectional mutation of MLT by \citet{Veach:1997}. There are a number of different ways in which this concept could be applied to light transport, e.g.\ a new perturbation that allows MMLT to transition between light paths of different length.

Finally, the detailed discussion of sample Jacobians presented in this paper could have applications outside of RJMLT acceptance probabilities. In a sense, the Jacobian determinant is a concise description of how sensitive the sampled path is to perturbations of the random numbers that sample it. This could lead to a novel adaptive step size control for primary sample space methods, which could \emph{locally} estimate the correct step size to reduce rippling effects and achieve optimal mixing rates.


\bibliographystyle{ACM-Reference-Format}
\bibliography{bibliography}

\end{document}